\documentclass[preprint2,10.5pt]{aastex}
\begin{document}

\title{\large{\rm{A CLUSTER OF CLASS I/f/II YSOs DISCOVERED NEAR THE CEPHEID SU CAS}}}
\author{D. Majaess$^1$, D. Turner$^1$, W. Gieren$^2$}
\affil{$^1$ Saint Mary's University, Halifax, Nova Scotia, Canada}
\affil{$^2$ Universidad de Concepci\'on, Concepci\'on, Chile.}
\email{dmajaess@cygnus.smu.ca}

\begin{abstract}
\textit{Preliminary} constraints are placed on a cluster of YSOs  (J2000 02:54:31.4 +69:20:32.5) discovered in the field of the classical Cepheid SU Cas.  WISE 3.4, 4.6, 12, and 22 $\mu m$ images reveal that the cluster deviates from spherical symmetry and exhibits an apparent diameter of $3 \times 6 \arcmin$.  SEDs constructed using 2MASS $K_s$ ($2.2 \mu m$) and WISE photometry indicate that 19 (36\%) class I, 21 (40\%) class f, and 13 (25\%) class II objects lie $r<3 \arcmin$ from the cluster center.   Conversely, 11 (18\%) class I, 13 (21\%) class f, and 37 (61\%) class II objects were detected for $r>3\arcmin$.   Approximately 50\% of the class I sources within $r<3 \arcmin$ were classified solely using WISE photometry owing to the absence of detections by 2MASS.
\end{abstract}

\section{{\rm \footnotesize INTRODUCTION}}
The latest generation of near and mid infrared surveys such as the VVV \citep[\textit{VISTA Variables in the Via Lactea},][]{mi10}, UKIDSS \citep[\textit{UKIRT Infrared Deep Sky Survey},][]{lu08}, GLIMPSE (\textit{Galactic Legacy Infrared Midplane Survey Extraordinaire}), and WISE \citep[\textit{Wide-field Infrared Survey Explorer},][]{wr10} surveys have fostered the discovery of countless stellar nurseries.  \citet{me05} discovered 92 star clusters using GLIMPSE data (3.6, 5.8, 8.0, $24 \mu m$), of which $\sim67$\% likely host a young stellar demographic. Similarly, a sizable fraction of the 96 clusters discovered by \citet{bo11} in the VVV survey ($YZJHK_s$) house heavily obscured young populations. Infrared photometry is particularly efficient at revealing embedded sources which are otherwise obscured from optical surveys since $A_{\lambda} \propto \lambda^{-\beta}$ \citep[e.g.,][]{ni09}.  The total extinction tied to photometry employed here is $A_{K_s}/ A_V\sim0.1$ for 2MASS $K_s$ ($\sim 2.2 \mu m$), and $A_{[3.4]}$:$A_{[4.5]}$:$A_{[12]}$:$A_{[22]} \sim0.05 A_{V}$ for WISE photometry \citep[e.g.,][]{fl07}.

\begin{figure*}
\begin{center}
\includegraphics[width=6.3cm]{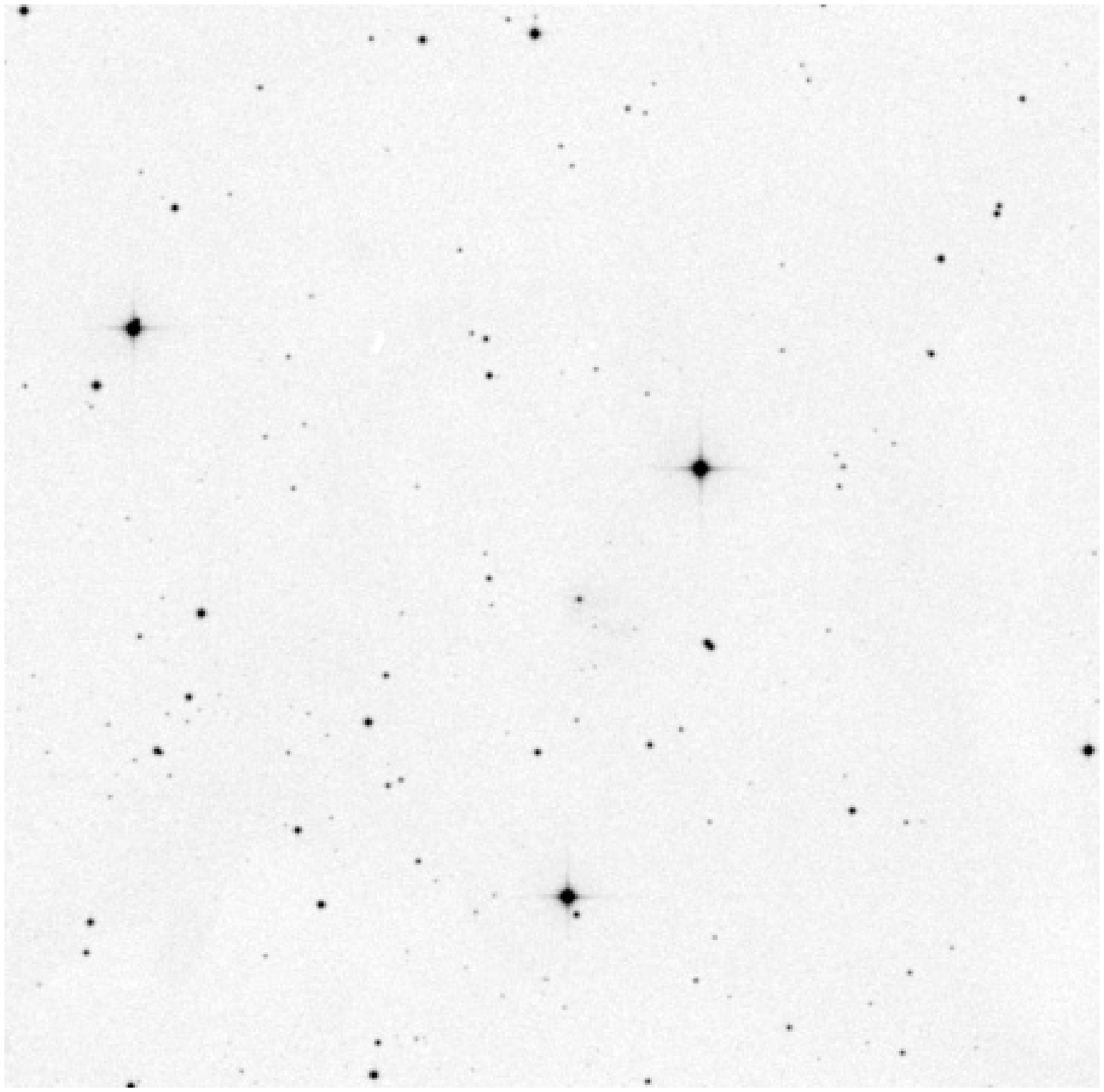}
\includegraphics[width=6.2cm]{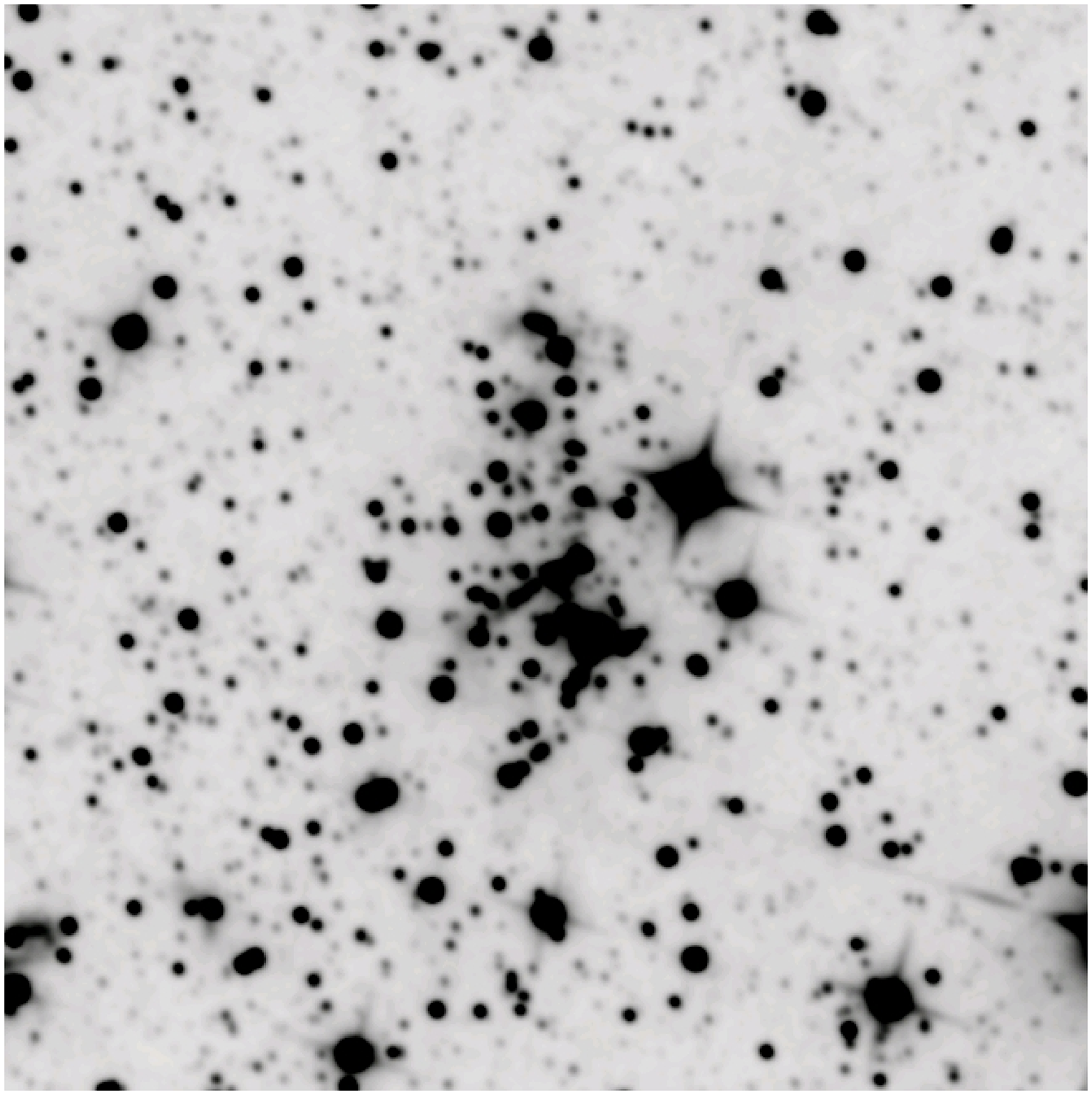}
\end{center}
\caption{\small{Left/right, optical DSS (\textit{Digitized Sky Survey}) and WISE 3.4 $\mu m$ images centered at J2000 02:54:31.4 +69:20:32.5.  The images span $\sim 15 \times 15 \arcmin$ and were aligned using the Aladin software environment \citep{bo00}.  The YSOs are absent from the optical image owing to significant interstellar and circumstellar extinction.}}
\label{fig-cl}
\end{figure*}

In this study, initial constraints are placed on young stellar objects (YSOs) discovered in the field of the $1.95^{\rm d}$ classical Cepheid SU Cas.  In \S \ref{s-cl}, optical (DSS) and $3.4 \mu m$ (WISE) images of the cluster are compared, and its apparent extent is determined from the latter; \S \ref{s-sed}, SEDs are constructed to classify the YSOs according to the prescription outlined by \citet{la87}, and to assess whether the YSO class varies as a function of radial distance from the cluster center; \S \ref{s-ccd}, $JHK_s$ color-color diagrams are constructed for cluster stars and an adjacent comparison field. 
\section{{\rm \footnotesize ANALYSIS}}
\subsection{{\rm \footnotesize UNVEILING THE CLUSTER}}
\label{s-cl}
The field of the classical Cepheid SU Cas hosts two clusters.  The first encompasses SU Cas, and \citet{te84} speculated that the Cepheid is a cluster member.  The connection is of particular importance since SU Cas is the shortest-period calibrator of the Galactic $VI_c$ Wesenheit function.  Such functions are employed to constrain $H_0$ and dark energy \citep{fm10}.  Observations obtained from the Abbey Ridge Observatory \citep{la08,ma08} for the parent cluster of SU Cas shall be described in a forthcoming study.  

The second cluster, detailed here, lies $\sim 30 \arcmin$ north east of SU Cas (Fig.~\ref{fig-cl}).  The cluster is absent from optical images of the region ($< 0.8 \mu m$).  A nearby YSO and an overdensity in 2MASS data were noted previously \citep[e.g.,][]{fr07,sk10}.  The 3.4, 4.6, 12, and $22 \mu m$ WISE images unambiguously reveal a cluster centered at J2000 02:54:31.4 +69:20:32.5.  

The cluster deviates from spherical symmetry, and exhibits an apparent diameter of $6 \arcmin$ in the north-south direction. The cluster's extent in the east-west direction is approximately half that value ($3 \arcmin$). 

\subsection{{\rm \footnotesize CLASSIFYING THE YSOs VIA SEDs}}
\label{s-sed}
The YSOs were classified by assessing the spectral energy distribution according to the approach of \citet{la87}.  The SED for each star was constructed in the canonical fashion:
\begin{equation}
\label{eqn-fd}
\log{(\lambda \times F_{\lambda})}=\alpha  \log{\lambda}+z \\
\end{equation}
\begin{equation}
\log{(\lambda \times F_{\lambda})}=\log{(\lambda \times 10^{m_\lambda /-2.5} F_{\lambda}(0)  kc/{\lambda}^2)}
\end{equation}
Where $\lambda$ is the wavelength of the passband (cm), $m_\lambda$ is the magnitude in that passband, $c$ is the speed of light (cm/s), $F_{\lambda}(0)$ is the zero-magnitude flux (Jy), $k$ is a conversion factor ($k=10^{-23}$ ergs cm$^{-2}$ s$^{-1}$ Hz$^{-1}$), and $\alpha$ is the slope of the function.   

\begin{figure}[!t]
\begin{center}
\includegraphics[width=7cm]{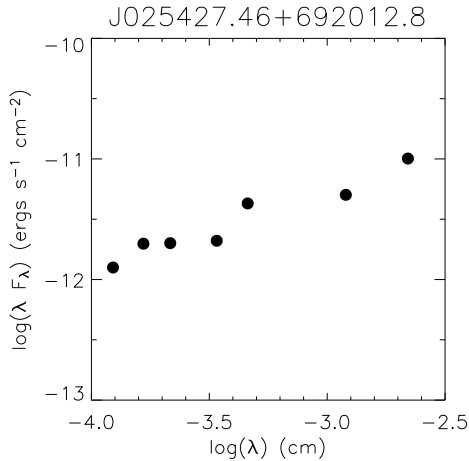}
\end{center}
\caption{\small{A SED for the YSO designated J025427.46+692012.8 constructed from 2MASS and WISE photometry.  The YSO is a class I source which lies $\sim 1 \arcmin$ from the cluster center.}}
\label{fig-sed}
\end{figure}

Class I, flat (hereafter class f), and class II YSOs exhibit $\alpha>0.3$, $-0.3<\alpha<0.3$, and $\alpha<-0.3$ accordingly \citep[see also][]{li11}.  The slope ($\alpha$) of equation (\ref{eqn-fd}) is determined from 2MASS $K_s$ and photometry extending redward of $2.2 \mu m$.  However, the wavelength dependence of extinction invariably biases the perceived slope.  Radiation emitted at 2MASS $K_s$ exhibits increased sensitivity to extinction than WISE passbands.  Simulations for J025427.46+692012.8 (class I) and J025435.68+692222.7
(class f) indicate that $\alpha$ varies according to: $\alpha_{(K_s+W)}^{'} \sim \alpha_{(K_s+W),0}-0.13 \times \delta A_{K_s}$.  That bias may be mitigated by determining $\alpha$ based solely on WISE photometry since the passbands display nearly equivalent extinction ratios ($A_{[3.4]}$:$A_{[4.5]}$:$A_{[12]}$:$A_{[22]} \sim0.5 A_{K_s}$).   In cases where $K_s$ was unavailable the determination of $\alpha$ resulted from WISE photometry. The correlation between the slope ($\alpha$) derived including and excluding 2MASS $K_s$ is described by:
$\alpha_{(K_s+W)} \sim 0.98 \times \alpha_{(W)} - 0.15$.

For 53 objects detected within $3 \arcmin$ of the cluster centre: 19 (36\%) are class I, 21 (40\%) are class f, and 13 (25\%) are class II objects.  Beyond $r>3\arcmin$ from the cluster center and the detection rates change to 11 (18\%) class I, 13 (61\%) class f, and 37 class II objects.  Within $r<3 \arcmin$ 10 (20\%) class I, 5 (10\%) class f, and 0 class II sources lack 2MASS photometry.  Beyond $>3 \arcmin$ a total of 9 (15\%) class I, 6 (10\%) class f, and 3 (5\%) class II sources were not catalogued by 2MASS.

\begin{deluxetable}{clcl}
\tablewidth{0pt}
\tabletypesize{\normalsize}
\tablecaption{YSO Candidates $<3 \arcmin$ from the Cluster Center}
\tablehead{\colhead{Designation} & \colhead{YSO Class} & \colhead{Designation} & \colhead{YSO Class}}
\startdata
J025426.03+691837.2	&	I	($K_s$+W)	&	J025428.46+691851.6	&	II	($K_s$+W)	\\
J025423.45+692011.6	&	I	($K_s$+W)	&	J025441.33+691911.1	&	II	($K_s$+W)	\\
J025423.60+692221.4	&	I	($K_s$+W)	&	J025432.89+692008.7	&	I	($K_s$+W)	\\
J025425.93+692251.4	&	f	($K_s$+W)	&	J025436.88+691911.3	&	f	($K_s$+W)	\\
J025437.43+692113.6	&	I	($K_s$+W)	&	J025434.48+692316.0	&	f	($K_s$+W)	\\
J025435.68+692222.7	&	f	($K_s$+W)	&	J025424.19+692051.9	&	f	($K_s$+W)	\\
J025439.36+691838.6	&	f	($K_s$+W)	&	J025437.54+691902.6	&	f	($K_s$+W)	\\
J025433.11+691753.3	&	f	($K_s$+W)	&	J025429.86+692221.9	&	I	($K_s$+W)	\\
J025425.62+691755.5	&	II	($K_s$+W)	&	J025437.03+692020.0	&	II	($K_s$+W)	\\
J025427.46+692012.8	&	I	($K_s$+W)	&	J025434.66+691933.9	&	II	($K_s$+W)	\\
J025430.67+692100.6	&	I	($K_s$+W)	&	J025438.85+691919.0	&	f	($K_s$+W)	\\
J025441.82+691945.9	&	f	($K_s$+W)	&	J025430.94+692034.8	&	I	(W)	\\
J025421.04+691932.9	&	II	($K_s$+W)	&	J025429.32+692000.0	&	f	(W)	\\
J025439.47+692244.1	&	f	($K_s$+W)	&	J025428.12+692057.3	&	I	(W)	\\
J025420.10+691954.7	&	f	($K_s$+W)	&	J025428.92+692122.1	&	I	(W)	\\
J025436.28+692154.2	&	I	($K_s$+W)	&	J025427.83+691953.5	&	I	(W)	\\
J025422.57+691948.7	&	f	($K_s$+W)	&	J025433.74+692139.4	&	I	(W)	\\
J025420.65+692122.5	&	II	($K_s$+W)	&	J025429.20+692141.1	&	I	(W)	\\
J025434.16+691824.0	&	f	($K_s$+W)	&	J025426.00+691951.3	&	I	(W)	\\
J025431.31+691929.6	&	II	($K_s$+W)	&	J025435.57+692139.0	&	f	(W)	\\
J025428.03+691801.1	&	II	($K_s$+W)	&	J025425.68+692121.3	&	f	(W)	\\
J025430.92+691919.1	&	II	($K_s$+W)	&	J025426.03+692207.7	&	f	(W)	\\
J025436.07+691844.0	&	II	($K_s$+W)	&	J025424.17+691915.3	&	f	(W)	\\
J025434.47+691835.3	&	II	($K_s$+W)	&	J025432.46+691805.3	&	I	(W)	\\
J025435.30+692246.0	&	f	($K_s$+W)	&	J025431.18+692311.2	&	I	(W)	\\
J025424.17+692136.5	&	f	($K_s$+W)	&	J025429.68+691742.5	&	I	(W)	\\
J025420.46+691919.5	&	II	($K_s$+W)	&		&			\\
\enddata
\label{table-1}
\end{deluxetable}

\begin{figure*}[!t]
\begin{center}
\includegraphics[width=10cm]{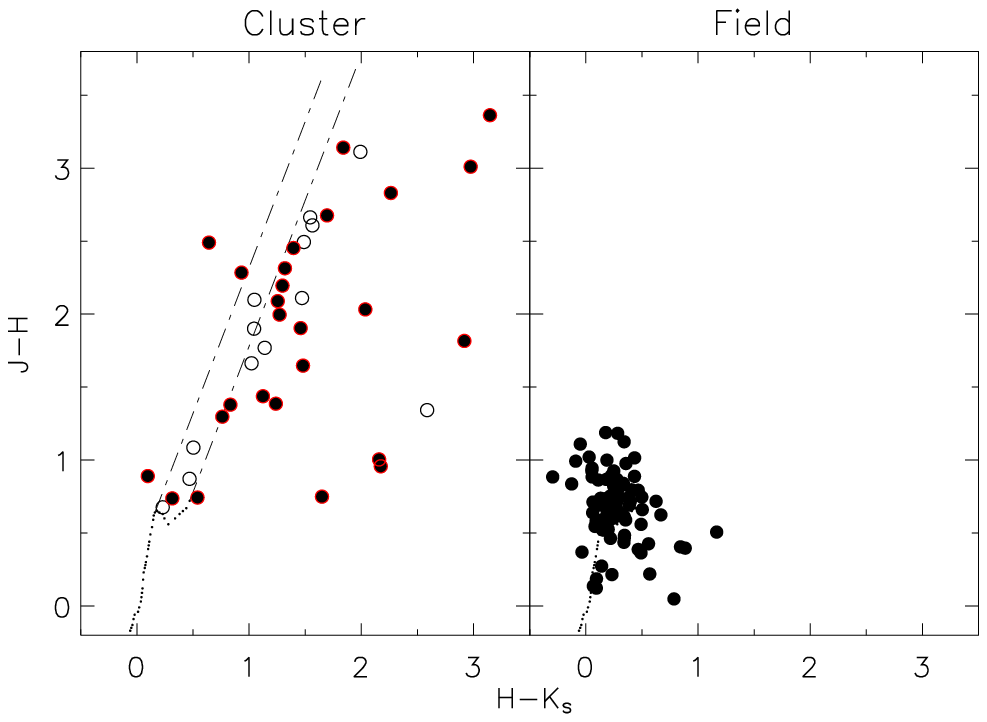}
\end{center}
\caption{\small{2MASS $JHK_s$ color-color diagrams for cluster stars (left panel) and a comparison field (right panel).  YSOs are absent from the latter, which is dominated by unreddened late-type field stars. The intrinsic $JHK_s$ main-sequence relation of \citet{sl08} is shown.  Reddened main-sequence stars typically lie within the region bounded by the parallel lines (dash-dot).  Left panel, open circles and red encircled black dots are  class II and class I/f candidates, accordingly.  Right panel, black dots denote late-type field stars.}}
\label{fig-ccd}
\end{figure*}

The aforementioned statistics reaffirm that the YSO class varies with radial distance from the cluster center.  The fraction of less-evolved objects decreases as a function of increasing radial distance. Admittedly, $\alpha$ may be biased by the effects of photometric contamination for stars near the crowded cluster core. 

\subsection{{\rm \footnotesize COLOR-COLOR DIAGRAM}}
\label{s-ccd}
$JHK_s$ color-color diagrams were constructed for the cluster stars and a comparison field $20\arcmin$ distant (Fig.~\ref{fig-ccd}).  The sequence of YSOs is absent from the comparison field.  The latter is dominated by unreddened late-type field stars according to the intrinsic $JHK_s$ relation of \citet{sl08}.  Conversely, several cluster members are heavily obscured ($A_V\sim 35$).  Indeed, radiation emitted from the 10 class I sources lacking $K_s$ photometry may suffer additional extinction.

The YSOs are highlighted on the color-color diagram by class (Fig.~\ref{fig-ccd}).  Identifications for the objects, which include their J2000 coordinates and classifications, are provided in Table~\ref{table-1}.  Discrete CO and HI emission  indicate that the cluster may be coincident with a complex foreground to SU Cas \citep{te84}, or beyond the Cepheid \citep[$d=418\pm12$ pc,][]{st11} at $d\la0.95$ kpc.  The latter sets a soft upper limit since most young clusters exhibit a height $Z$ less than that implied (note that $\ell,b\sim133.5,9\degr$).   Additional research is required.

\section{{\rm \footnotesize CONCLUSION}}
Countless YSOs were discovered at J2000 coordinates 02:54:31.4 +69:20:32.5 (Fig.~\ref{fig-cl}), which is $30 \arcmin$ north east of the classical Cepheid SU Cas.  The cluster appears non-symmetric and displays an apparent diameter of $6 \arcmin$ (Fig.~\ref{fig-cl}).  SEDs constructed using 2MASS and WISE photometry indicate that the cluster hosts 19 (36\%) class I, 21 (40\%) class f, and 13 (25\%) class II objects (Fig.~\ref{fig-sed}).  At a distance beyond 3$\arcmin$ from the cluster center the sampling changes markedly to 11 (18\%) class I, 13 (20\%) class f, and 37 (60\%) class II objects.  The statistics reaffirm that the YSO class exhibits a radial dependence.  10 class I sources lacking 2MASS $K_s$ photometry were classified via SEDs tied to WISE observations.  The slope $\alpha$ (equation \ref{eqn-fd}) inferred from that approach appears less biased by the wavelength dependence of extinction since WISE passbands exhibit nearly equivalent extinction ratios (as compared to $K_s$).  The extreme extinction obscures many of the cluster's YSOs ($A_V\sim 35$) beyond optical detection (Fig.~\ref{fig-ccd}).  

The latest generation of infrared surveys (e.g., WISE, GLIMPSE, etc.) shall continue to foster the detection and characterization of new stellar clusters, as reaffirmed here.  Additional research, which includes spectroscopic follow-up of the cluster and broader region, is required.

\subsection*{{\rm \scriptsize ACKNOWLEDGEMENTS}}
\scriptsize{DM is grateful to the following individuals and consortia whose efforts lie at the foundation of the research: 2MASS, WISE, J. Gizis, E. Paunzen (WebDA), W. Dias (DAML), CDS, arXiv, and NASA ADS.  WG is grateful for support from the Chilean Center for Astrophysics FONDAP 15010003 and the BASAL Centro de Astrofisica y Tecnologias Afines (CATA) PFB-06/2007.  This publication makes use of data products from the Wide-field Infrared Survey Explorer, which is a joint project of the University of California, Los Angeles, and the Jet Propulsion Laboratory/California Institute of Technology, funded by NASA.}

\end{document}